# Rederivation of Nernst's equation without any additional assumptions

Shanh Su, Jincan Chen[1]

Department of Physics, Xiamen University, Xiamen 361005, People's Republic of China

It is found that without any additional assumptions, Nernst's equation can be re-deduced from the experimental data obtained from the thermodynamic systems at ultra-low temperatures, and consequently, the physical content included by Nernst's equation should not be referred to as Nernst's postulate or Nernst's theorem. It should be called the Nernst statement. This discovery will play an important role in improving the theoretical framework of thermodynamics. It can effectively prevent some artificial assumptions into the third law of thermodynamics, making which be a true reflection of the objective world. It solves the awkward problem caused by using a thermodynamic theorem as the core contents of a thermodynamic law for over a hundred years.

[1]Email: jcchen@xmu.edu.cn

The Nernst theorem [1] is the essence of the third law of thermodynamics and occupies an important position in the thermodynamic theory [2-5]. However, in the derivation of the Nernst theorem, some additional assumptions were introduced [2, 6-8]. As early as 1920, Lewis and Gibson [9] pointed out that "the third law has been verified experimentally, but we also see some a priori reasons for the existence of such a law." Although the contents and statements of the third law of thermodynamics has long been a controversial issue [10-23], the derivation mode of the Nernst theorem in textbooks has rarely been suspected since Nernst's time [2, 6-8, 24]. Recently, it was found [25] that the derivation of the Nernst equation does not need those additional assumptions appearing in textbooks. Below, we will elaborate on this problem so that Nernst's equation is restored to its original rightful glory.

Before Nernst, many chemists had studied the chemical reactions of thermodynamic systems at ultra-low temperatures, measured the changes $\Delta H$ and $\Delta G$ of the enthalpy and Gibbs function of thermodynamic systems at isothermal and isobaric pressures, and found that both $\Delta H$ and $\Delta G$ get closer and closer as the temperature is dropped, as shown by the solid curves in Fig.1 [1, 2, 7, 24].

According to the first law of thermodynamics, the relation between $\Delta H$ and $\Delta G$ of a general thermodynamic system in the isothermal process is given by

$$\Delta G = \Delta H - T\Delta S, \tag{1}$$

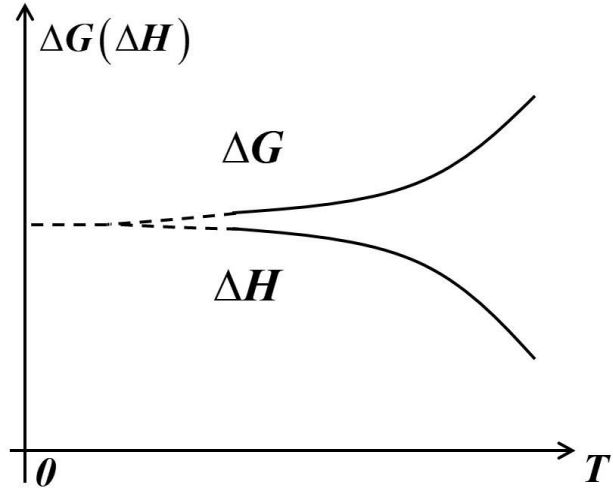

Fig.1. The schematic diagram of $\Delta H$ and $\Delta G$ varying with $T$.

where $T$ and $S$ are the temperature and entropy of the system, respectively. It is seen from the dashed curves in Fig.1 that when the experimental results are extrapolated to absolute zero temperature, one may get $\Delta G = \Delta H$. Nernst [1] published a diagram (reproduced as Fig. 2 in Ref. [24]), illustrating that the two curves of $\Delta H$ and $\Delta G$ varying with $T$ might have come together in almost any fashion and still have satisfied the condition that $\Delta H$ and $\Delta G$ are equal at absolute zero, which gives no information on the shapes of the curves at absolute zero temperature [24].

Moreover, it is assumed that when $T \to 0$, $\Delta S$ is bounded [2, 6, 7] so that Eq. (1) can be expressed as

$$\lim_{T \to 0} \frac{\Delta H - \Delta G}{T} = \lim_{T \to 0} [\frac{\partial \Delta H}{\partial T} - \frac{\partial \Delta G}{\partial T}] = \lim_{T \to 0} \Delta S . \qquad (2)$$

As described in textbooks, there are two ways to derive the Nernst theorem from Eq. (2). One way is to suppose that both $\Delta H \sim T$ and $\Delta G \sim T$ curves are tangent to each other and have the same slope as $T \to 0$ [6, 8], i.e.,

$$\lim_{T \to 0} \frac{\partial \Delta H}{\partial T} = \lim_{T \to 0} \frac{\partial \Delta G}{\partial T}. \tag{3}$$

This was what Nernst asserted in 1906 [26]: for solids and liquids, the slopes of $\Delta H$ and $\Delta G$ were asymptotic as $T \to 0$. He did not include ideal gases in his statement, because the law of ideal gas loses definition at absolute zero [24]. It means that $\Delta H$ and $\Delta G$ are not only equal to each other at the absolute zero temperature, but also that their values coincide completely in the immediate vicinity of this point [24].

From Eqs. (2) and (3), one can derive the Nernst theorem, i.e.,

$$\lim_{T \to 0} (\Delta S)_T = 0. \tag{4}$$

The other way is to suppose directly that the Nernst theorem, Eq. (4), is true [2, 7], and consequently, Eq. (3) is obtained, which means that both $\Delta H \sim T$ and $\Delta G \sim T$ curves have the same slope to be equal to zero as $T \to 0$.

In the above derivation, two additional assumptions are introduced, so that the Nernst theorem is also called as the Nernst postulate [2, 21, 27]. This makes Nernst's theorem less bright than it should have.

If directly using the experimental data of both $\Delta H$ and $\Delta G$ obtained from the chemical reactions of thermodynamic systems at ultra-low temperatures, one can easily calculate the values of $y \equiv |\Delta H - \Delta G|$ and obtain the curves of $y$ varying with $T$, as indicated in Fig.2. It should be pointed out that like Fig.1, Fig.2 is also a schematic diagram, which does not affect to display the re-derivation of the Nernst equation [i.e., Eq. (4)].

Using the data in Fig.2 and computational simulation, one can find that $y_i$ $(i = 1, 2, ...)$ is a function of $T$. For different thermodynamic systems, $y_i$ may have different forms. However, a large number of experimental results have shown that for general thermodynamic systems at ultra-low temperatures, $y_i$ has one common feature: the decrease of $y_i$ with temperature is faster than that of $T$ [28]. The simplest form of $y_i$ is $y_i \propto T^i$, where $i > 1$.

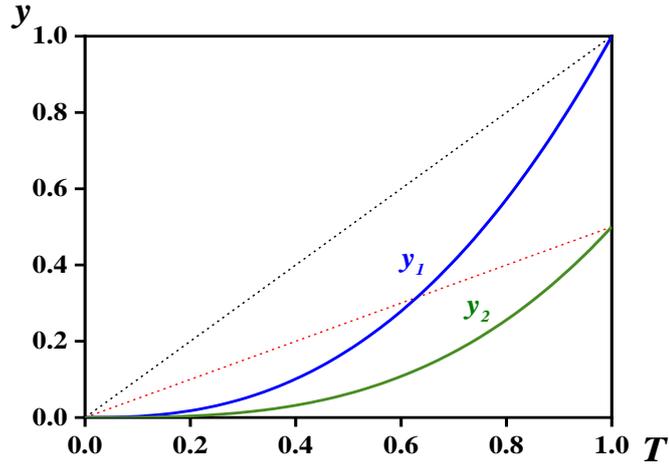

Fig.2. The schematic diagram of $y$ varying with $T$.

Substituting $y_i$ into Eq. (1) and extrapolating it to absolute zero temperature, one can directly derive

$$\lim_{T \to 0} \frac{|\Delta H - \Delta G|}{T} = \lim_{T \to 0} \frac{y_i}{T} = \lim_{T \to 0} |(\Delta S)_T| = 0 = \lim_{T \to 0} (\Delta S)_T. \tag{5}$$

The last equation in Eq. (5) is exactly the Nernst Equation.

For a new thermodynamic system, using the experimental data of $\Delta H$ and $\Delta G$ at ultra-low temperatures and the method mentioned above, one can simulate the curve equation of $|\Delta H - \Delta G|$ varying with $T$ and derive Eq. (5).

It is very important to note the fact that in the re-derivation process of Eq. (5), no additional assumptions are required. This is different from the derivation of Nernst's theorem in textbooks. Thus, the physical content included in Eq. (5) should not be referred to as Nernst's postulate or Nernst's theorem and may be called the Nernst statement, i.e., the entropy change associated with any isothermal reversible process of a thermodynamic system approaches zero as the temperature approaches absolute zero. Like the Clausius statement or the Kelvin-Planck statement [4, 7, 29] as one of the important statements of the second law of thermodynamics, the Nernst statement can be used as one of the main statements of the third law of thermodynamics so that the core contents of the third law of thermodynamics are a true reflection of the objective world without any artificial additional assumptions.